\documentclass[12pt,cites,graphicx]{article}

\usepackage{epsfig}

\title{Characterization of the complex ion dynamics in lithium silicate glasses via computer simulations}

\author{Andreas Heuer,  Magnus Kunow, Michael Vogel, and Radha D. Banhatti,\\[3mm]
Institute of Physical Chemistry, Schlossplatz 4/7, D-48149
M\"unster \\[3mm] and Sonderforschungsbereich 458\\[3mm]
andheuer@uni-muenster.de\\
kunow@uni-muenster.de\\mivogel@uni-muenster.de\\banhatt@uni-muenster.de}

\begin{document}

\maketitle

\renewcommand{\thefootnote}{\fnsymbol{footnote}}

\noindent

We present results of molecular dynamics simulations on lithium
metasilicate over a broad range of temperatures for which the
silicate network is frozen in but the lithium ions can still be
equilibrated. The lithium dynamics is studied via the analysis of
different correlation functions. The activation energy for the
lithium mobility agrees very well with experimental data. The
correlation of the dynamics of adjacent ions is weak. At low
temperatures the dynamics can be separated into local vibrational
dynamics and hopping events between adjacent lithium sites.  The
derivative of the mean square displacement displays several
characteristic time regimes. They can be directly mapped onto
respective frequency regimes for the conductivity. In particular
it is possible to identify time regimes dominated by localized
dynamics and long-range dynamics, respectively.  The question of
time-temperature superposition is discussed for the mean square
displacement and the incoherent scattering function.

\section{Introduction}

\label{intro} The dynamics of ions in amorphous materials is very
complex, as indicated, e.g., by the strong frequency dependence of
the conductivity \cite{Funke,Dyre,Roling,Roling2}. This frequency
dependence directly reflects the presence of back- and
forthdynamics of the individual ions. The complexity of the ion
dynamics is due to the simultaneous action of the time-dependent
Coulomb interaction with the other mobile ions and the
time-independent interaction with the spatially disordered and
basically immobile network.  Interestingly, the frequency
dependence of the conductivity is very similar when comparing
different materials \cite{Funke}. Even simulations of disordered
hopping models display a similar frequency dependence; see, e.g.,
\cite{Dyre,Dieterich,Maass}.

Typically the frequency-dependence of the conductivity
$\sigma(\nu)$ displays several characteristic time regimes. For a
sodium silicate system \cite{Wong} one observes, e.g., a microsopic
regime for $\nu
> 10^{13}$ Hz, a scaling regime
$\sigma(\nu) \propto \nu^2$ for  $10^{13}$ Hz $ > \nu > 10^{11}$
Hz, and finally for  $\nu < 10^{11}$ Hz a continuous decrease of
$\sigma(\nu)$ with decreasing $\nu$. In a double-logarithmic
representation at low temperatures the apparent exponent decreases
from one to zero until the d.c. plateau is reached, i.e.
$\sigma(\nu) = \sigma_{d.c.}$. A possible interpretation of this
behavior relates these frequency regimes to local vibrations,
stochastic localized dynamics, and jump dynamics, respectively
\cite{Conny}. An important observation for the frequency-dependent
conductivity $\sigma(\nu)$ is the validity of the time-temperature
superposition principle. For most materials it can be expressed by
the Summerfield scaling for which $\sigma(\nu)/
\sigma_{dc}$ is only a function of $\nu/(T \sigma_{dc})$
\cite{Summerfield}.

It is important to relate the frequency dependence of
$\sigma(\nu)$ to the real space dynamics of the ions. In this way
one may get interesting  information about the nature of the ion
dynamics. This can be done by expressing $\sigma(\nu)$ in terms of
the mean square displacement $\langle r^2(t) \rangle$.
Conceptually, both observables differ in one important aspect.
While the conductivity contains effects of multi-particle
correlations, the mean square displacement is a single-particle
quantity. Experimentally, the correlation is expressed by the
Haven ratio  \cite{Haven,Isard} which in general can be frequency
dependent (denoted $H_R(\nu)$). The experiments indicate that the
correlation among different ions is small and furthermore can be
approximated by a value $H_R$. $1/H_R$ is a measure for the number
of ions which are significantly correlated. Then one can relate
$\sigma(\nu)$ to $\langle r^2(t) \rangle$  via linear response
theory \cite{Funke,Kubo}
\begin{equation}
\label{eqnur2}
 \sigma(\nu) =  \frac{q^2}{6 H_R \rho k_B T}
\int_0^\infty dt \, (d/dt)w(t) \cos(2\pi \nu t)
\end{equation}
where $q$ denotes the charge, $\rho$ the density of the mobile
ions, and $H_R$ the Haven ratio. The function $w(t)$ is defined as
\begin{equation}
w(t)= (d/dt) \langle r^2(t) \rangle.
\end{equation}

Analysis of spatial aspects of dynamical processes is one of the
strongholds of computer simulations. Here it is of particular help
that detailed information is available on a microscopic level.
Conceptually, one can proceed in two steps.  In a first step one
can study $w(t)$. On the basis of Eq.\ref{eqnur2} it has been
readily shown that for a significant dispersion of the
conductivity the frequency-dependence of $\sigma(\nu)$ and the
time-dependence of $w(t)$ are basically mirror images of each
other when identifying $t$ with 1/$\pi^2 \nu$ \cite{Funke}.
Therefore it is the quantity $w(t)$ which is of uttermost
importance for the direct relation to conductivity experiments
since one may hope to recover the three regimes of $\sigma(\nu)$
also in $w(t)$. As a consequence of Eq.1 a dispersion in
$\sigma(\nu)$ is equivalent to a time dependence of $w(t)$ and a
scaling relation $w(t) \propto t^{-\alpha}$ transfers to
$\sigma(\nu) \propto \nu^\alpha$.
 In a second step one can analyse the dynamics in more
detail to elucidate the underlying physical processes in the
different time regimes. Although interesting details may be seen
from the analysis of individual trajectories a sound understanding
of the dynamics should be based on the study of appropriate
correlation functions, yielding the {\it average} dynamical
behavior.

In this paper we present computer simulations for lithium silicate
following this general approach.  As an example we take the lithium
dynamics in (Li$_2$O)(SiO$_2$) \cite{Habasaki4,Habasaki6,Cormack2}.
In contrast to most earlier work on this system the present
computer generation allows one to study the lithium dynamics in the
low-temperature dispersive regime in equilibrium conditions.
Similar systems like sodium silicate systems are also of current
interest for numerical studies
\cite{Horbach1,Horbach2,Kob1,Horbach3,Oviedo,Cormack,
Kieffer1,Kieffer2,Smith,Smith2}. The outline of this paper is as
follows. In Section 2 we describe the technical aspects of the
simulation and discuss the numerical tools.  The results of our
simulations as well as their interpretation are presented in
Section 3. We close with a discussion and a summary in Section 4.

\section{Simulation}

The potential energy of the lithium silicate system can be written
as the sum of a Buckingham and Coulomb pair potential
\begin{equation}
U_{ij}(r) = \frac{q_i q_j e^2}{r} - \frac{C_{ij}}{r^6} + A_{ij}
\exp(-B_{ij} r ).
\end{equation}
The pair indices i,j characterize the various ion pairs (Li-Li,
Li-Si, etc.). The individual potential parameters are listed in
our previous work \cite{Banhatti}. They are based on the work of
Habasaki et al.
\cite{Habasaki4,Habasaki6,Habasaki1,Habasaki2,Habasaki5} and have
been obtained from ab-initio calculations. Effective charges
$q_{Si} = 2.4, q_{Li} = 0.87$, and $q_{O} = 1.38$ have been
chosen. This choice fulfills charge neutrality of the total
system. Periodic boundary conditions have been used. Simulations
during the production runs were performed in the NVE ensemble. The
elementary timestep of our molecular dynamics simulations was 2
fs, the density $\rho = 2.34$ g cm$^{-3}$, taken from experimental
room temperature data \cite{Doweidar}. This corresponds to
pressures of the order of 1 GPa at the lowest temperatures of our
simulation. The total number of atoms in our system is 1152. For
generating the trajectories we have used the programm MOLDY,
supplied by K. Refson \cite{Refson}. In our longest simulation run
the system has been propagated for 20 ns. For our simulations we
have started from configurations at T = 1500 K where ions as well
as network could be equilibrated. Equilibration of the total
system at this temperature requires a simulation run of ca. 4 ns.
We note that the computer glass transition as observed on the time
scale of 10 ns is approximately 1100 K (see network diffusion data
in \cite{Banhatti}). In the present simulation we have chosen one
temperature slightly above $T_g$ (T = 1240 K) and four
temperatures below $T_g$ (T = 980 K, 750 K, 700 K, and 640 K). For
these four temperatures we have paid attention to equilibrate the
lithium subsystem before recording our observables. The network
only performs local fluctuations with an average radius much
smaller than one Angstrom. Interestingly, despite this fact we
observed a slight temperature drift as a result of the continuous
relaxation of the network for temperatures below the glass
transition.

\section{Results}

\subsection{Trajectories}

In Fig.1 some representative trajectories of individual lithium
ions at $T = 750$ K are shown. For these ions the dynamics can be
described as a series of local vibrations and jumps between
adjacent ionic sites. Similar features have been observed in
several previous simulations on alkali silicates, see e.g. Refs.
\cite{Kieffer1,Kieffer2,Smith}. The three ions, displayed  in
Fig.1, behave somewhat differently. Whereas ion (1) does not leave
its intitial site at all, ion (2) performs several back- and
forthjumps between two adjacent sites, and ion (3) follows a
random-walk like path between adjacent sites. The variety of
different motional patterns directly shows that it is important to
use appropriate statistical observables to identify the average
type of behavior of the lithium dynamics.  This is the subject of
the remainder of the paper.

\subsection{Structure}

In Fig.2 we display the partial Li-Li structure factor $g(r)$ for
several temperatures. One can clearly see that $g(r)$ shows a
sharp nearest neighbor peak at $r_{nn} \approx 2.6$ \AA.
Interestingly, the width of the nearest-neighbor peak of $g(r)$
only slightly changes in the temperature range between $T = 750$ K
and $T = 1240$ K. Here we have included also a very high
temperature $T = 4000$ K from our previous simulations
\cite{Banhatti}. Only for the higher temperature the nearest
neighbor peak broadens. This shows that the low-temperature width
of the nearest-neighbor peak is mainly due to the disorder of the
network and not due to thermal fluctuations in crystalline-like
ionic sites.

To study the degree of clustering of the lithium ions we have
calculated the function
\begin{equation}
h(r) = \frac{\int_0^r ds g(s) 4 \pi s^2}{4\pi r^3/3}.
\end{equation}
$h(r)$ counts the number of particles around a central particle in
a sphere of radius $r$ relative  to the number one would have for
a random distribution of particles. Thus the value of $h(r)$ is a
direct measure whether or not a particle is surrounded by more
$(h(r) > 1$) or less $(h(r) < 1$) particles than expected for a
random distribution. The first neighbor shell, as obtained from
the first minimum of $g(r)$ ranges up to $r \approx 4.4$ \AA.
Therefore the value $h(r = 4.4$ \AA) yields information whether
the number of particles in the nearest neighbor shell is smaller
or larger than expected for a statistical distribution. The
function $h(r)$ is displayed in Fig.3. It turns out that for low
temperatures $h(r = 4.4$ \AA) $\approx 0.9$. This indicates that
the repulsion effect of the ions due to their mutual Coulomb
interaction dominates the structure of the lithium subsystem. For
$T = 4000$ K the distribution is essentially statistical, as
expected for a dominance of entropic contributions. It is known
experimentally from the study of alkali silicates
\cite{Eckert,Greaves} that for smaller alkali concentration the
alkali ions start to form clusters and indeed we observe in so far
unpublished work that for (Li$_2$O)9(SiO$_2$) $h(r = 4.4 $
\AA$)\approx 1.5$.

\subsection{Dynamics}

At first we would like to present results for the mean square
displacement $\langle r^2(t) \rangle$ at different temperatures;
see Fig.4. For all temperatures there exists a significant
subdiffusive regime which is more pronounced for the lower
temperatures. Note that at much higher temperatures the
subdiffusive regime disappears and one observes a direct transition
from a ballistic short-time regime $\langle r^2(t)
\rangle \propto t^2$ to a diffusive long-time regime $\langle
r^2(t) \rangle \propto t$ \cite{Banhatti}. The  diffusion constant
$D$ can be directly estimated from the long-time regime via $D
= \langle r^2(t) \rangle /(6 t)$.

Next we present the temperature dependence of the diffusion
constant $D(T)$. The results are shown in Fig.5. We have included
high-temperature data points as already presented in our previous
work \cite{Banhatti}. The dynamics is characterized  by a single
activation energy $E_a = 0.58$ eV in the temperature range between
$T = 6000$ K and $T = 640$ K.  The computer glass transition $T_g
\approx 1100$ K does not show up in the temperature dependence of
the diffusion constant because already above this temperature the
ion and network dynamics are strongly decoupled.

For a comparison of the activation energy with experimental
results one should consider constant pressure rather than constant
density simulations. In order to check how much the
above-mentioned value of $E_a$ is modified  we took the pressure
of the T=1500 K run and checked how much the diffusion constant at
T=6000 K is modified when using constant pressure rather than
constant density. It turned out that the diffusion constant
increased by a factor of 1.6. This corresponds to an increase of
the activation energy by 0.08 eV, i.e. $\tilde{E}_a = 0.66$ eV.
This value compares very well with the experimental value of 0.59
eV as obtained from conductivity experiments \cite{Bunde}.

 In Fig.6 we show that all curves can be
superimposed on each other by substituting $t$ by $t^* =
(D(T)/D_0)t$ ( The constant $D_0$ is chosen as $D(T=640$ K)). Thus
one may conclude that in the scaling regime, starting beyond the
ballistic regime, $\langle r^2(t^*) \rangle$ fulfills the
time-temperature superposition principle. The relation to the
experimental findings will be discussed in the next Section.

From a theoretical perspective the incoherent scattering function
$S(q,t)$ is an important observable to characterize the dynamics.
It is defined as
\begin{equation}
S(q,t) = \langle \cos (\vec{q} (\vec{r}(t) - \vec{r}(0))) \rangle.
\end{equation}
For isotropic systems it only depends on the absolute value $q$ of
the wave vector $\vec{q}$.
 In particular for $q = q_{max} = 2\pi/r_{nn} $ (in our case $q_{max} = 2.4$ \AA$^{-1}$)
  one gets
information about the dynamical processes on the lengthscale of
nearest neighbor lithium distances. For glass forming liquids one
can formulate a theory which characterizes the time and wave vector
dependence of $S(q,t)$ for a broad range of temperatures
\cite{Goetze}. In Fig.7 one can find $S(q_{max},t)$ at different
temperatures. The decay time is an appropriate measure for the time
scale on which an ion effectively moves to the next site.

For glass-forming liquids it is predicted from theory and
confirmed by simulation that $S(q,t)$ follows the time-temperature
superposition principle in the long-time regime
\cite{Goetze,KobAndersen}.  Interestingly, also for the present
case of the lithium dynamics time-temperature superposition is
observed. This can be seen from Fig.7 where at all temperatures
$S(q_{max},t)$ is fitted to the Kohlrausch-Williams-Watts function
$ A \exp(-(t/\tau)^\beta)$ with $\beta = 0.45$. Actually, the
temperature dependence of $\tau$ is also activated (not shown). In
the range of temperatures analysed in this work the variation of
$\tau$ is four times larger than the variation of the diffusion
constant. Thus translates into a higher activation energy of 0.70
eV rather than 0.58 eV as obtained for the diffusion constant. A
possible origin of this descrepancy is discussed in the next
Section.

The primary goal of this paper is to elucidate the nature of the
frequency dependence of the conductivity. As already discussed the
conductivity and the mean square displacement are directly related
if the Haven ratio $H_R$ is constant. This is particularly the case
if $H_R$ is close to one, i.e. dynamical inter-ion correlations are
small. In recent work we have introduced a quantitity $N_{coop}(t)$
which is a measure of cooperativity on a time scale $t$
\cite{Doliwa00}. It is defined as
\begin{equation}
N_{coop}(t) \equiv \frac{\sum_{ij} X_i(t) X_j(t)}{\sum_i X_i(t)
X_i(t)}
\end{equation}
where $X_i(t)$ is a dynamic property of ion $i$ on the time scale
$t$ which on average is zero, i.e. $\sum_i X_i(t) = 0$. Here we
choose
\begin{equation}
\label{eqx}
 X_i(t) \equiv  (\vec{r}_i(t) - \vec{r}_i(0))^2 -
\langle r^2(t) \rangle.
\end{equation}
For the practical implementation of this approach it is essential
that $N_{coop}$ is calculated for subsystems. The technical
details of this analysis can be found in \cite{Doliwa00}. The
results are shown in Fig.8. For a better comparison of the
different temperatures we have plotted $N_{coop}$ against the
normalized time $t^*$. As expected $N_{coop} (t^*)$ approaches one
on very short and very long time scales since the different ions
behave independently. For all temperatures the maximum occurs at
approximately the same scaled time $t^* \approx 500 ps$. This time
corresponds to the regime where the transition from the strongly
subdiffusive to the diffusive regime occurs. For the lower
temperatures there are approximately 4 particles which are
dynamically correlated. In contrast to glass-forming systems there
is only a minor temperature dependence of $N_{coop}$.  A further
difference for glass-forming liquids is that the maximum value of
$N_{coop}$ can be larger than 20 close to the glass transition
\cite{Doliwa00}. Therefore the present values of $N_{coop}$ are
indeed small.

Is $N_{coop}$ related to the Haven ratio? On a qualitative level
both quantities express the dynamic cooperativity. In the most
simple case exactly $M \ll N $ particles behave identically but
all these subsets of $M$ particles behave in an uncorrelated
manner. Then one simply has
\begin{equation}
\label{eqncoop}
 N_{coop} = 1/H_R = M.
\end{equation}
On a more quantitative level we start from the zero-frequeny limit
of the Haven-ratio, thereby slightly extending the arguments given
in \cite{Doliwa00}. This limit is given by
\begin{equation}
\label{defh}
 H_R = \frac{\sum_i \int_0^\infty dt \, \langle v_i(0)
v_i(t) \rangle}{\sum_{i,j} \int_0^\infty dt \, \langle v_i(0)
v_j(t) \rangle}.
\end{equation}
These terms can be further rewritten in order to obtain the
connection with $N_{coop}$. Due to time-reversal symmetry of the
molecular dynamics one has  $\langle v_i(0) v_j(t) \rangle$ =
$\langle v_i(0) v_j(-t) \rangle$. We define $t_m$ as a time far in
the diffusive long-time regime, i.e. much longer as the time $t_d$
for which the velocity-correlations have basically decayed to
zero. Then one can write
\begin{equation}
\int_0^\infty dt \, \langle v_i(0) v_j(t) \rangle = (1/2)
\int_{-t_m}^{t_m} dt \, \langle v_i(0) v_j(t) \rangle.
\end{equation}
For $-t_m + t_d < \tau < t_m - t_d$ one can substitute $v_i(0)$ by
$v_i(\tau)$. Thus, for $t_m \rightarrow \infty$ one can write
\begin{equation}
\int_{-t_m}^{t_m} dt \, \langle v_i(0) v_j(t) \rangle = (1/2t_m)
\int_{-t_m}^{t_m} d\tau \, \int_{-t_m}^{t_m} dt \, \langle
v_i(\tau) v_j(t) \rangle.
\end{equation}
For stationary processes one can substitute the integration limits
by 0 and 2$t_m$, respectively. Rewriting Eq.\ref{defh} in terms of
the above relations and replacing $ \int_{0}^{2t_m} dt \, v_i(t)$
by $r_i(2t_m) - r_i(0)$ one ends up by $H_R = 1/N_{coop}(t
\rightarrow \infty)$ when choosing $X_i(t) = \vec{r}_i(t) -
\vec{r}_i(0)$. Thus Eq.\ref{eqncoop} also holds more generally.
Unfortunately, for the present data the statistical error was very
large with this choice of $X_i(t)$ and much smaller with the
choice in Eq.\ref{eqx}. However, for the hard sphere system
analysed in \cite{Doliwa00} it turned out that to a very good
approximation the maximum of $N_{coop}(t)$ with time for a
quantity involving the absolute value of the shift vectors (like
the one used above) is very close to the long-time limit of
$N_{coop}$ for the choice $X_i(t) = \vec{r}_i(t) - \vec{r}_i(0) $.
Thus the maximum value of $N_{coop}$ in Fig.8 is indeed a good
measure for the inverse zero-frequency Haven ratio. Of course, a
more detailed analysis of this aspect would be desirable. The
resulting low-temperature estimate $H_R \approx 1/4$ is in
agreement with typical experimental values \cite{Isard}. Finally
we would like to mention that even this modest cooperativity has
been stressed as a key feature to understand the long-time
diffusion \cite{Habasaki8}.

For a closer discussion of $w(t)$ (and thus of $\sigma(\nu)$) it
will turn out to be helpful to use additional information about the
real-space dynamics. For this purpose we have determined the
van-Hove self-correlation function $G_s(r,t)$ for the lithium ions
\cite{Hansen}
\begin{equation}
G_s(r,t) \equiv \langle \delta(r - |(\vec{r}(t) -
\vec{r}(0))|)\rangle.
\end{equation}
It denotes the probability that a lithium ion moves a distance $r$
during the time $t$. The results are shown in Fig.9. One can
clearly see that at short times the dynamics of all ions is
confined to a small r-range. Of course, for longer times the ions
may explore larger regions of the system. The presence of the
first well-resolved peak around $d_0 \approx 2.6$ \AA \, shows
that hopping processes with jump length $d_0$ are a relevant
feature of the dynamics. To a good approximation $d_0$ is
identical to the r-value $r_{nn}$ of the first nearest-neighbor
peak of the partial structure factor $g(r)$.  Also the other peaks
of the partial structure factor show up in $G_s(r,t)$. Thus one
may conclude that the network provides well-defined lithium sites
which are separated by barriers larger as compared to $k_B T$. The
lithium ions basically jump between adjacent sites.  The term {\it
jump} implies that the time scale it takes to cross the saddle
 is much shorter than the time scale the ions
fluctuate within the individual sites. This conclusion is in
qualitative agreement with previous simulations
\cite{Cormack2,Smith,Cormack1,Habasaki7,Balasubramanian} and also
agrees with the appearance of the trajectories in Fig.1.

The
 first minimum of $G_s(r,t)$ around
$r_{min} \approx 1.5$ \AA\ can be used to distinguish local
vibrational dynamics ($|\Delta r(t) | < 1.5$ \AA) and long-range
 dynamics ($|\Delta r(t) | > 1.5$ \AA). Of course, there
may be individual ions which display local fluctuations with a
length scale larger than 1.5 \AA. On average, however, this value
of 1.5 \AA \, turns out as a significant length scale in the
above-mentioned sense. This scenario of jumps between well-defined
potential wells implies that the ions explore the individual sites
for rather long times before they leave the site on a time scale
approximately given by the decay time of $S(q_{max},t)$. In order
to analyse the intra-site dynamics somewhat closer we have
determined the mean square displacement $\langle r^2(t)
\rangle_{local}$ of those particles which at time $t$ have moved
less than $1.5$ \AA. This curve is shown at all four temperatures
of our study in Fig.10. For times larger than 0.2 ps $\langle
r^2(t) \rangle_{local}$ starts to become constant. This means that
already on this short time scale the lithium ions have largely
explored the local potential well. The increase of $\langle r^2(t)
\rangle_{local}$ for $T = 1240$ K at long times is related to the
fact that the network starts to relax since we are above its glass
transition temperature. Note that the temperature dependence of
this local exploration time is very weak. A careful inspection
shows that for the two lower temperatures there is further gradual
increase until $t \approx 10$ ps ($T = 750$ K) and $t \approx 100$
ps ($T = 640$ K), respectively.  This observation indicates the
presence of a substructure of the different wells. At low
temperatures the exploration of the initial well might require the
crossing of some lower saddles. This would explain the slower
approach to the final plateau value as compared to the dynamics in
a smooth local potential.

For long times all curves approach a constant plateau value
$\langle r^2(t \rightarrow \infty) \rangle_{local}$. An exception
is $T = 1240$ K where we take the value around $t = 10$ ps. In
Fig.11 this plateau value is plotted against temperature. To a
good approximation $\langle r^2(t \rightarrow \infty)
\rangle_{local}$ is proportional to
 temperature. This is expected for the dynamics in a
harmonic potential. Thus the local dynamics can be characterized by
an effective harmonic potential. These local fluctuations are
related to the Debye-Waller factor \cite{Angell}.

Now we are in a position to discuss $w(t)$. The lengthscale
$r_{min} = 1.5$ \AA  \, as derived from the van Hove self
correlation function allows us to separate local and long-range
motion and was the basis to study $\langle r^2_{local}(t)
\rangle$. In analogy to $w(t)$ we define
\begin{equation}
w_{local}(t) \equiv (d/dt) \langle r^2_{local}(t) \rangle.
\end{equation}
It is shown in Fig.12 at different temperatures. The temperature
dependence of $w_{local}$ is rather weak. For $t > 20$ fs it
roughly scales like $t^{-2}$.

 In Fig.13
$w(t)$ is shown for several temperatures. Three different time
regimes can be distinguished. (i) For $t < 20 fs$ there is the
ballistic regime, i.e. $w(t)
\propto t$. (ii) For $20fs < t < 1 ps$ there is a strong decay of
$w(t)$.  (iii) For $t
> 1 ps$ $w(t)$ decays much slower until it becomes constant for
large times. Whereas for $T = 1240$ K and $T = 980 $K this final
decay is very weak it becomes significantly larger for lower
temperatures. For $T = 640$ the decay spans even  more than a
decade.

It is instructive to compare this with the typical experimental
$\sigma(\nu)$ data for alkali silicates, as already discussed in
the Introduction. The frequency regimes there have a one-to-one
correspondence to the characteristic time regimes of $w(t)$. (i)
The vibrational modes for $\nu
> 10^{13}$ Hz correspond to the ballistic regime in the time domain. Since
$1/(\pi^2 \nu) \approx 10 fs$ this agrees very well with the
corresponding time regime of $w(t)$ (ii) As mentioned before the
$\sigma(\nu) \propto \nu^2$ behavior for $ 10^{13} $ Hz$
>\nu
> 10^{11}$ Hz can be attributed to stochastic dynamics in a harmonic potential
 \cite{Conny}. For the
time dependence of $w(t)$ of an overdamped vibration one expects
$w(t) \propto \exp(-kt)$ with some decay constant $k$ \cite{Doi}.
For a complicated system like an amorphous ion conductor a
distribution of decay constants should be present, resulting in a
somewhat modified decay characteristics, but still yielding
$\sigma(\nu) \propto \nu^2$ for $\nu < 1/k_{max}$ as obtained via
Fourier transformation. Here $k_{max}$ denotes the upper cutoff of
the distribution of decay constants. In any event, one expects a
very strong decay in agreement with the behavior in $w(t)$ for
times $20 fs < t < 1 ps$. (iii) The more gradual decrease of
$\sigma(\nu)$ for $\nu < 10^{11}$ is also recovered in our
simulated $w(t)$ curves for the two lower temperatures. For $T \ge
950$ K $w(t)$ is roughly constant for $t
> 1 ps$. For the sodium silicate system the dispersion of
$\sigma(\nu)$ as roughly expressed by $\sigma (\nu \approx 10^{11}
Hz)/\sigma_{dc}$ is about $10^7$ at room temperature. It becomes
smaller with increasing temperature and vanishes at around $T
\approx 900$ K.

The postulation of a crossover from vibrational-type to long-range
dynamics around $\nu = 10^{11}$ Hz, as derived from conductivity
spectra, is consistent with physical intuition. There is, however,
no strict derivation. Here simulations may give additional
information because the microscopic nature of dynamics is known in
detail. In particular, we have identified the contribution
$w_{local}(t)$ which arises from local vibrational dynamics, see
Fig.12.  As expected its value is close to $w(t)$ for short times.
The long-range contribution to $w(t)$ can be estimated as $w(t) -
w_{local}(t)$. This is included in Fig.13 for the  two lowest
temperatures. For these temperatures one can clearly see that the
dispersion for $t > 1 ps$ is mainly due to long-range motion. For
$T = 640$ K the dispersion, i.e. $w(t=1 ps)/w(t \rightarrow
\infty)$, is around 20.

\section{Discussion and Summary}

The analysis as presented above allows us to improve our
understanding of the frequency dependence of the conductivity.
Here it is helpful that the time-dependence of $w(t)$ (derivative
of the mean square displacement) is strongly related to the
frequency dependence of $\sigma(\nu)$.  As outlined above several
findings like the activation energy or the Haven ratio are in good
agreement with the corresponding results of experiments. In
particular it turns out that only for temperatures below 900 K the
system displays dispersive dynamics due to long-range dynamics.
Thus only simulations in the nanosecond regime can cope with the
effect of back- and forthjumps as observed in the experiment. At
first view this result disagrees with the fact that even at $T =
1240$ K the mean square displacement (see Fig.4) is not diffusive
but shows a subdiffusive regime. Our analysis of $w(t)$, however,
has revealed that the subdiffusive regime for this high
temperature is related to local vibrational dynamics. Therefore it
only contributes to the high-frequency regime of $\sigma(\nu)$,
reflecting vibrational properties.

Long-range dynamics prevails for $t
> 1$ ps or, correspondingly, $\nu < 10^{11}$ Hz. This observation
seems to contradict the conclusion in Ref.\cite{Ngai} that the
nearly constant loss and thus the dynamics for frequencies much
lower than $10^{11}$ Hz is due to anharmonic vibrational dynamics
rather than jump dynamics. A priori, however, it is not clear
whether long-range dynamics automatically implies  jump-like
motion and further work is necessary to clarify this point.

As shown in Fig.13 the subtraction of $w_{local}(t)$ from $w(t)$
yields a plateau-like region for short times. This is in agreement
with recent work on amorphous  0.5 Ag$_2$S - 0.5 GeS$_2$
\cite{Conny}. Of course, for experimental data it is not possible
to determine the vibrational contribution individually. For some
materials, however, it can be estimated reliably \cite{Conny}. The
interpretation of $w(t) - w_{local}(t)$ as the contribution of
jumps between discrete sites is definitely justified for long
times. In general it is justified as long as the vibrational and
the jump dynamics are uncorrelated to each other. Intuitively one
would expect that this is the case for time scales which are
significantly longer than the time it takes for the ion to
transfer between two adjacent sites. Two important questions
remain. What is this crossing time? What is the physical
interpretation of $w(t) - w_{local}(t)$ for shorter times than the
crossing time? Work along this line is in progress.

Another important aspect is related to the time-temperature
superposition principle. The scaling, observed in Fig.6, is
consistent with the Summerfield scaling as can be easily checked
with the help of  Eq.\ref{eqnur2}. As is evident from
Eq.\ref{eqnur2} measurement of $\sigma(\nu)$ is not sufficient to
obtain an absolute value of the mean square displacement.  From
measurements of $\epsilon^\prime (\infty)$, however, it becomes
possible to get information about the absolute value $c(T)$ of
$\langle r^2(t) \rangle$  for short times (or high frequencies,
respectively)
 where only the local fluctuations contribute \cite{Roling,Roling2,Happe}. In
agreement with our results (see Fig.11) one finds to a good
approximation
 $c(T) \propto T$ \cite{Happe}. A more detailed comparison of the
 experimental scaling results  and our findings for the
 mean square displacement is not possible so far since our scaling
 regime is still very limited and simulations at still lower
 temperatures had to be taken into account.

Finally, we would briefly comment on the different activation
energies of the conductivity and the incoherent scattering
function. In glass-forming systems at low temperatures a similar
effect is seen when comparing the rotational correlation function
(which basically behaves like the translational incoherent
scattering function) and the diffusion constant. As explained in
Ref.\cite{Sillescu} such an effect may be related to a broadening
of the waiting time distribution for decreasing temperature.  The
diffusion constant is mainly dominated by the fast ions. In
contrast, decay of the incoherent scattering function (with time
scale $\tau$) requires that also the slow ions move. Thus an
increase of heterogeneity would increase the product of the
diffusion constant and the relaxation time $\tau$. On a
qualitative level this is seen in our simulations. Although the
heterogeneities indeed increase with decreasing temperature
\cite{Heuer02} we do not know whether this is the only explanation
for the different activations energies. In any event, this
question is accessible experimentally since via multidimensional
NMR observables can be measured which are believed to behave like
the incoherent scattering function \cite{Vogel}.

The increase of heterogeneity as indirectly seen by the different
activation energies implies that in a very strict sense the
dynamics at lower temperatures is not a simple time-scaled version
of the dynamics at higher temperatures. Thus we have to conclude
that the mean square displacement and the incoherent scattering
function, which both can be interpreted as reduced representations
of the dynamics, are not sensitive to these deviations from
time-temperature superposition. Actually, a similar observation
has been already reported for a glass-forming hard sphere system
for which the non-gaussian parameter strongly depends on density
although the mean square displacement displays time-density
superposition \cite{Doliwa99}.

Several important questions remain to be answered. How do the
observed features correlate with the local network structure (see
Refs.\cite{Kob1,Horbach3} for interesting features observed
recently)? To which degree do the observed features change with
concentration? What can one learn about the nature of correlated
back- and forthjumps? For the latter question it turns out to be
very helpful to study appropriate three-time correlation functions
\cite{Heuer02}.

 Of course, the results of several experiments
as well as previous simulations already contain part of the
answers. However, since computer simulations are always dealing
with temperatures much higher than experiments one has to be very
careful to extract information from simulations which is relevant
for experiments. Therefore the advent of faster computers and thus
the ability to simulate at sufficiently low temperatures will
render computer simulations of ion conductors a fruitful field for
the future.

\section{Acknowledgement}

In this work we have greatly benefited from helpful discussions
with C. Cramer, K. Funke, J. Horbach, and B. Roling. We
acknowledge the support by the DFG (SFB 458). Furthermore we would
like to acknowledge K. Refson for supplying the MOLDY software
package.

\clearpage


\clearpage

\begin{list}{}{\leftmargin 2cm \labelwidth 1.5cm \labelsep 0.5cm}

\item[\bf Fig. 1]
Trajectories of three different lithium ions at $T = 750$ K as
projected on a plane. The box length is 9 \AA.

\item[\bf Fig. 2]
The partial structure factor $g(r)$ for the lithium ions at
different temperatures.

\item[\bf Fig. 3]
The integrated partial structure factor $h(r)$ for the lithium
ions at different temperatures.

\item[\bf Fig. 4] The mean square displacement $\langle r^2(t)
\rangle$ for lithium at different temperatures.

\item[\bf Fig. 5]
The temperature dependence of the diffusion constant of Li$_2$O
SiO$_2$. The data marked as circles have been taken from
Ref.\cite{Banhatti}.

\item[\bf Fig. 6] The same as in Fig.4.
The individual times have been scaled in order to show the
time-temperature superposition.

\item[\bf Fig. 7]
The time dependence of the incoherent scattering function $S(q,t)$
for $q = q_{max} = 2\pi/d_0$ at different temperatures. The broken
lines correspond to a fit with the KWW function $A
\exp(-(t/\tau)^\beta)$ with $\beta = 0.45$.

\item[\bf Fig. 8]
The time dependence of dynamic cooperativity as expressed by
$N_{coop}(t)$ for different temperatures.

\item[\bf Fig. 9] The self part of the van Hove correlation function $G_s(r,t)$
for different times $t$ at $T=750$ K.

\item[\bf Fig. 10] The local mean square displacement $\langle r^2(t)
\rangle_{local}$ which averages over all displacements less than
$r_{min} = 1.5$ \AA.

\item[\bf Fig. 11] The temperature
dependence of $\langle r^2(t \rightarrow \infty) \rangle_{local}$.

\item[\bf Fig. 12] $w_{local}(t)$ at different temperatures.

\item[\bf Fig. 13] $w(t)$ at different temperatures. For the two
lowest temperatures also $w(t) - w_{local}(t)$ is displayed.

\end{list}

\begin{figure}
\centerline{\epsfxsize=8cm\epsffile{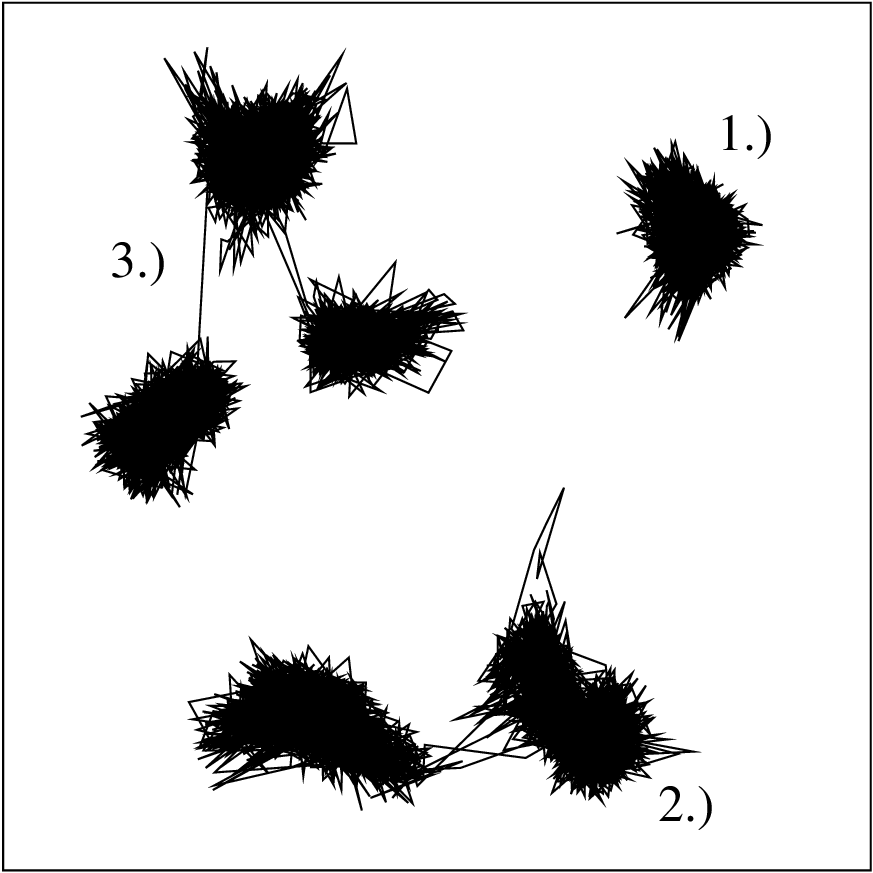}}
\centerline{\bf Fig.\ 1}
\end{figure}
\hfill

\begin{figure}
\centerline{\epsfxsize=11.5cm\epsffile{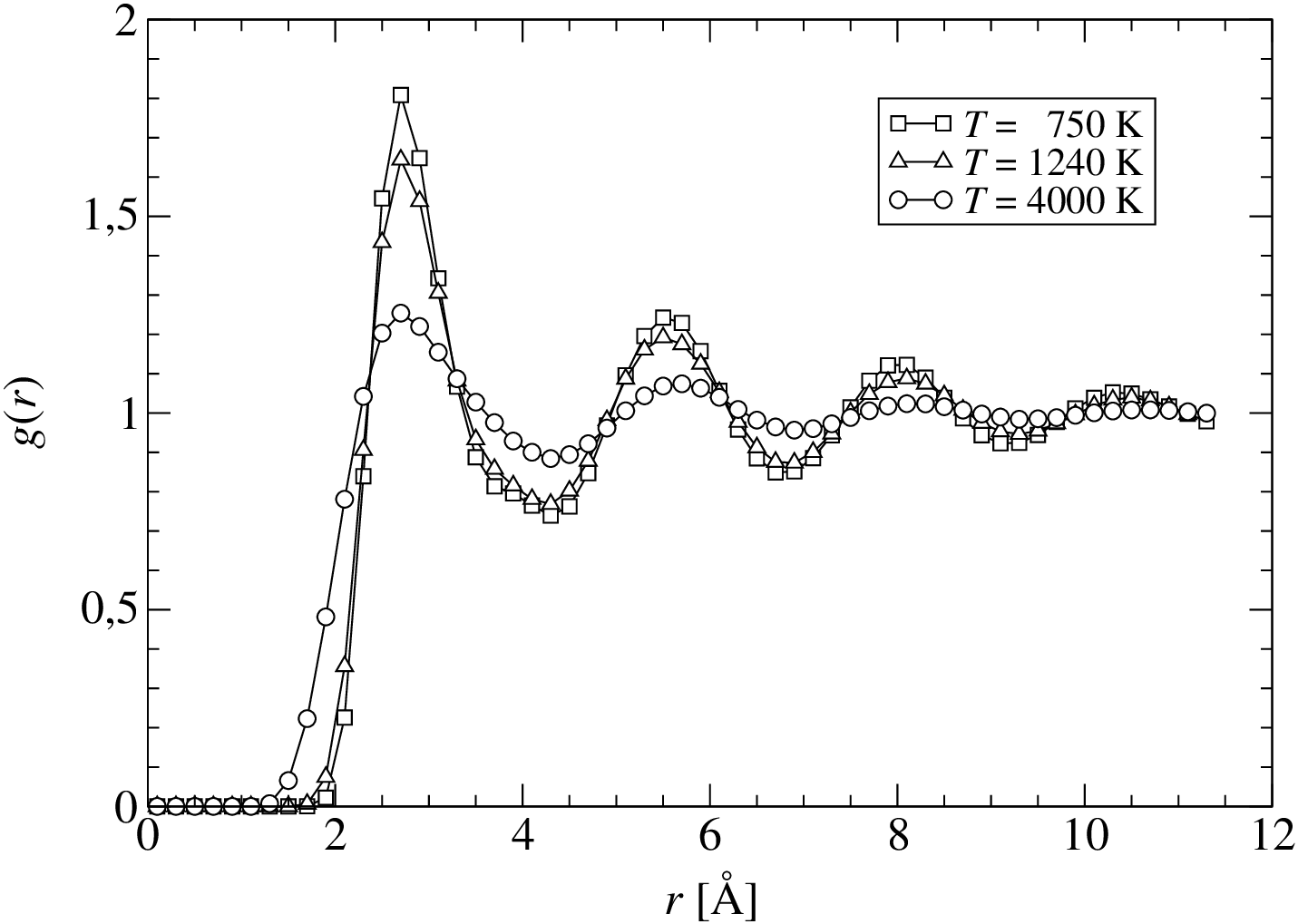}}
\centerline{\bf Fig.\ 2}
\end{figure}

\begin{figure}
\centerline{\epsfxsize=11.5cm\epsffile{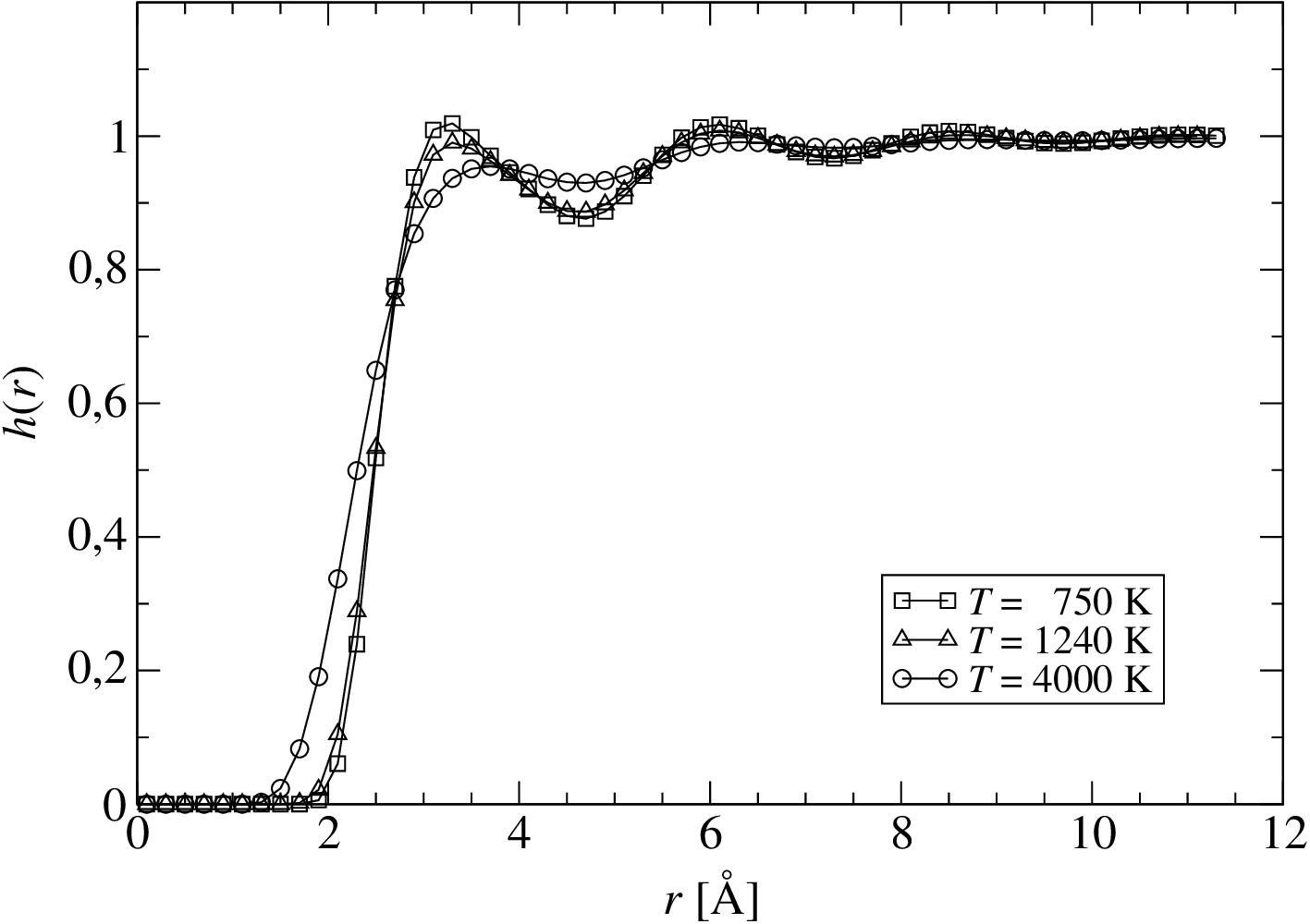}}
\centerline{\bf Fig.\ 3}
\end{figure}
\hfill

\begin{figure}
\centerline{\epsfxsize=11.5cm\epsffile{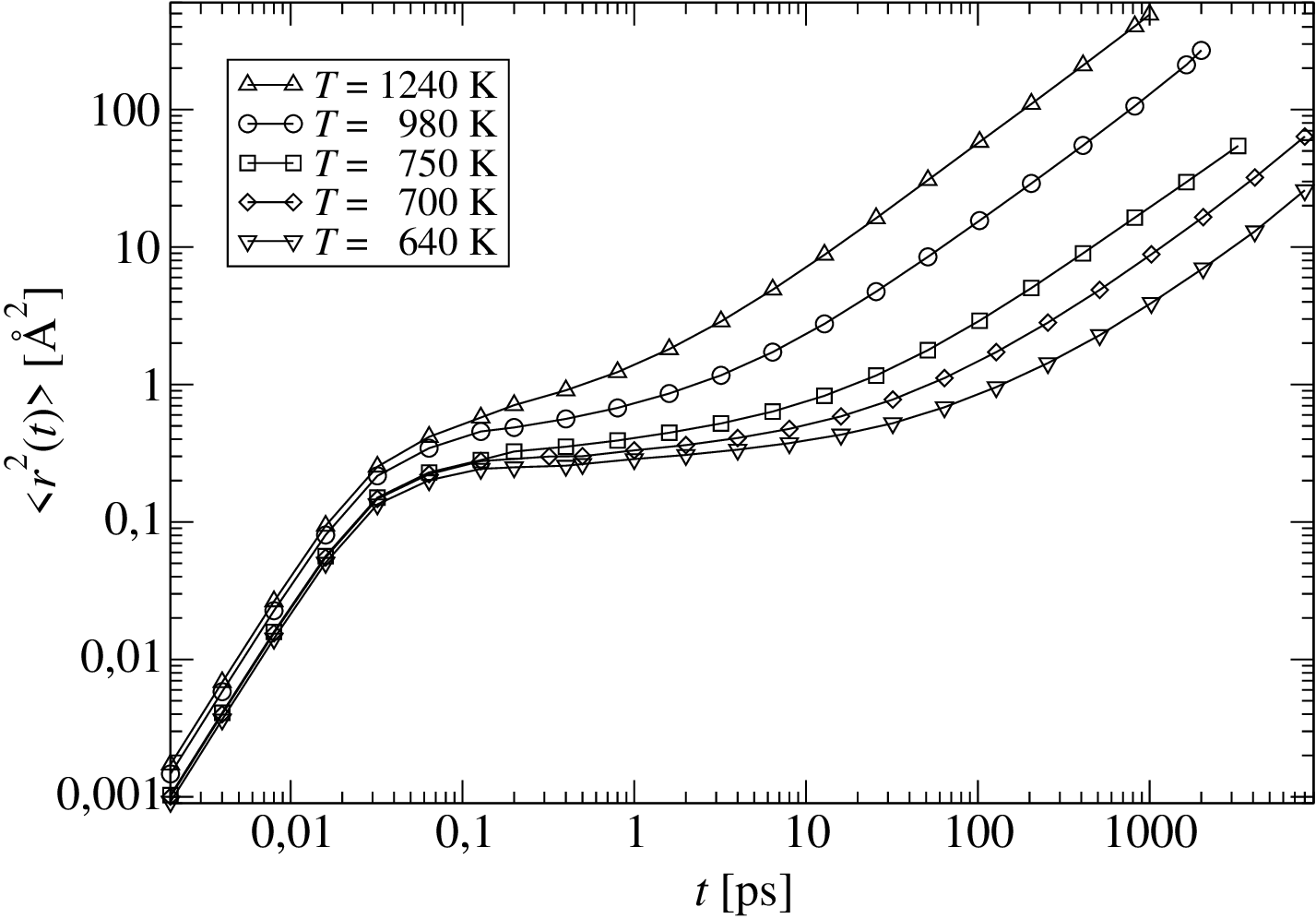}}
\centerline{\bf Fig.\ 4}
\end{figure}

\begin{figure}
\centerline{\epsfxsize=11.5cm\epsffile{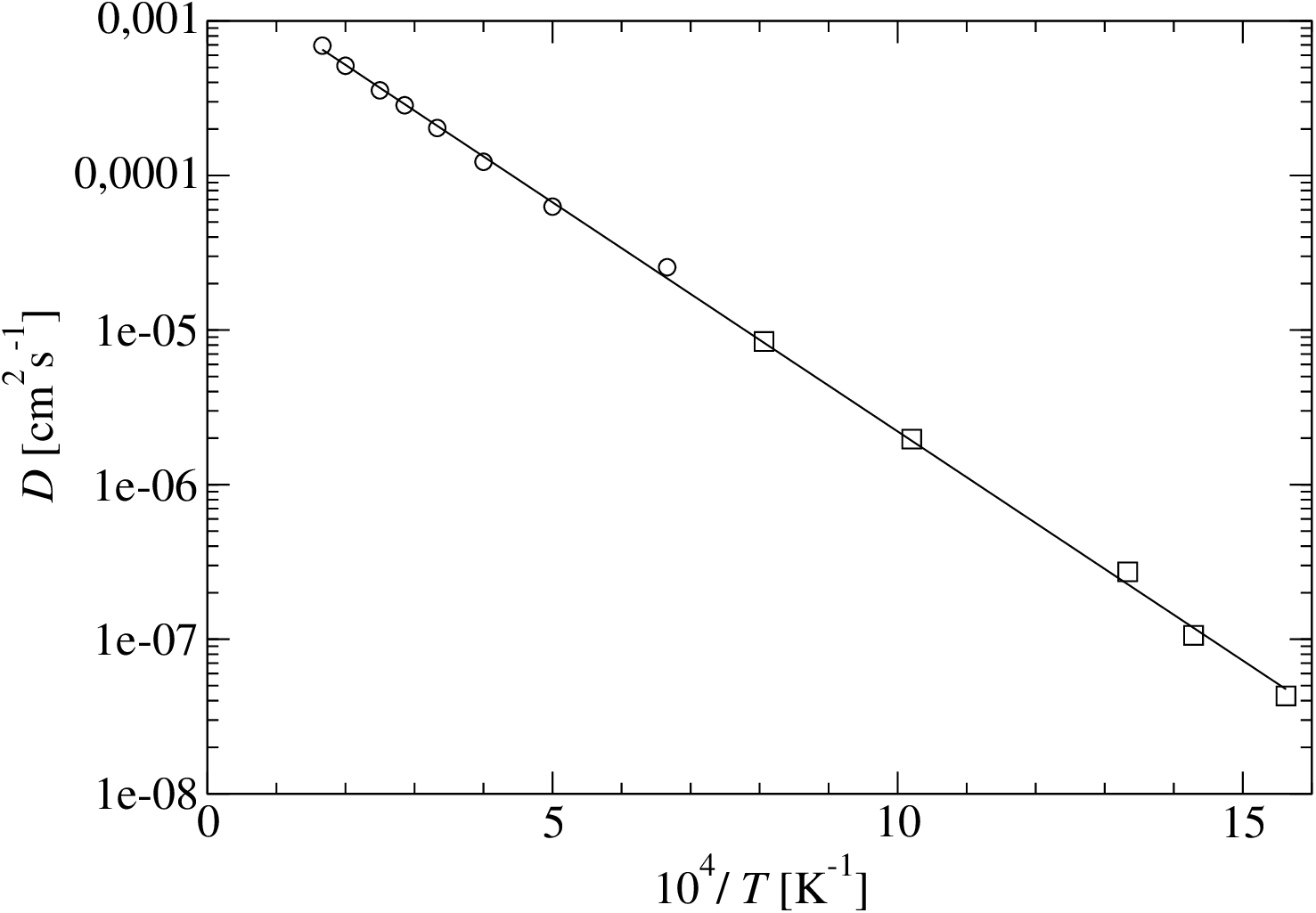}}
\centerline{\bf Fig.\ 5}
\end{figure}
\hfill

\begin{figure}
\centerline{\epsfxsize=11.5cm\epsffile{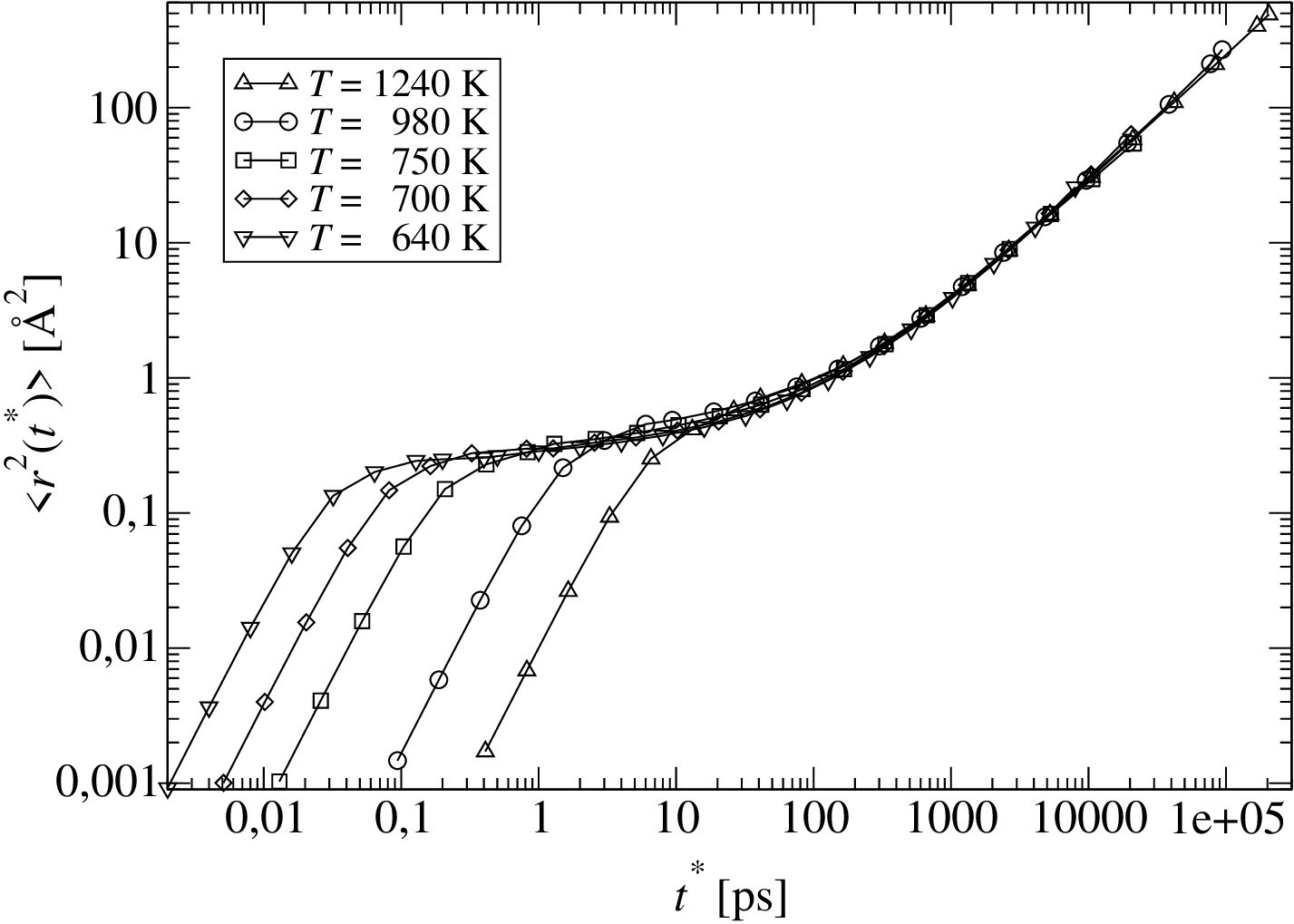}}
\centerline{\bf Fig.\ 6}
\end{figure}

\begin{figure}
\centerline{\epsfxsize=11.5cm\epsffile{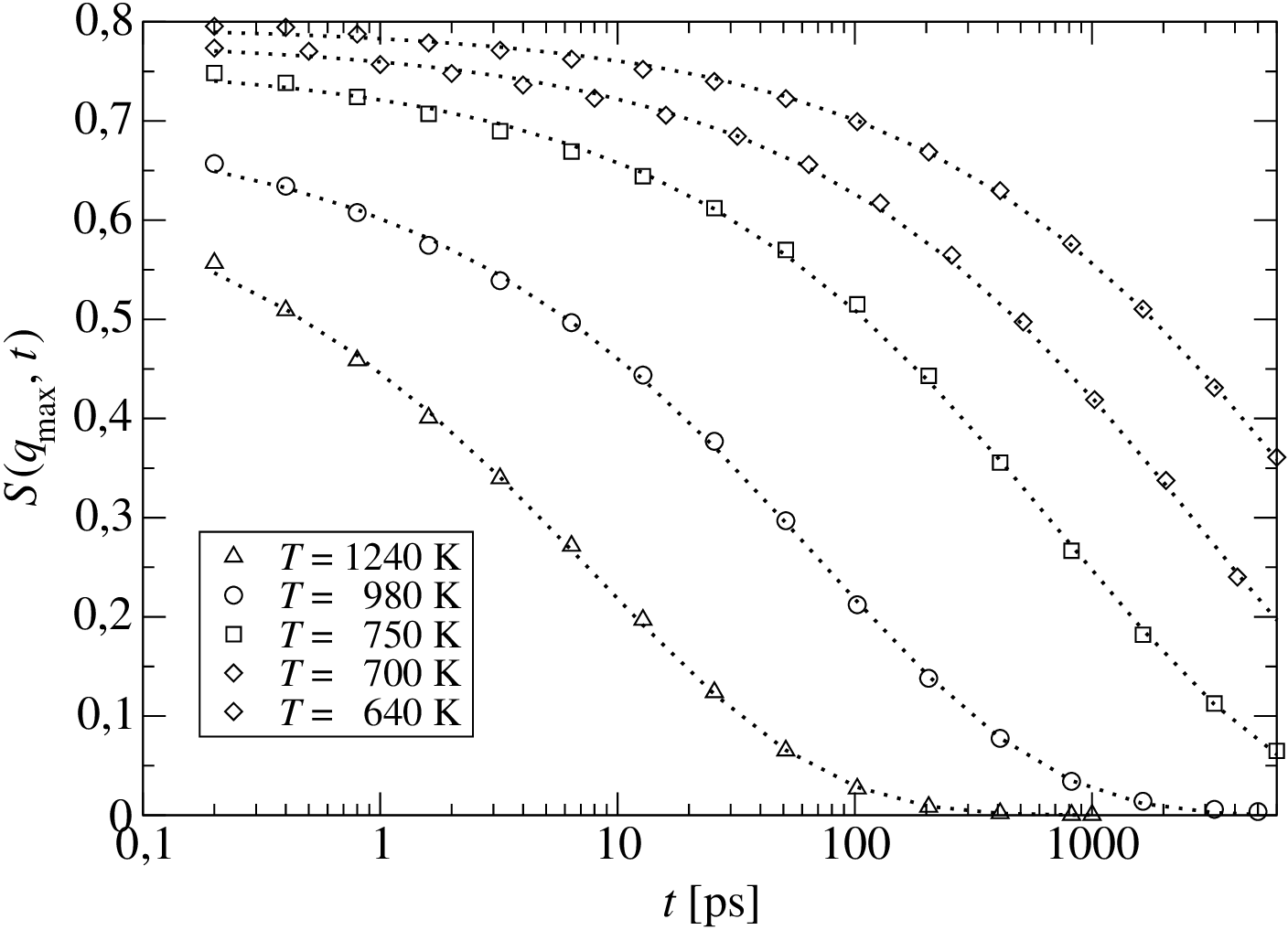}}
\centerline{\bf Fig.\ 7}
\end{figure}
\hfill

\begin{figure}
\centerline{\epsfxsize=11.5cm\epsffile{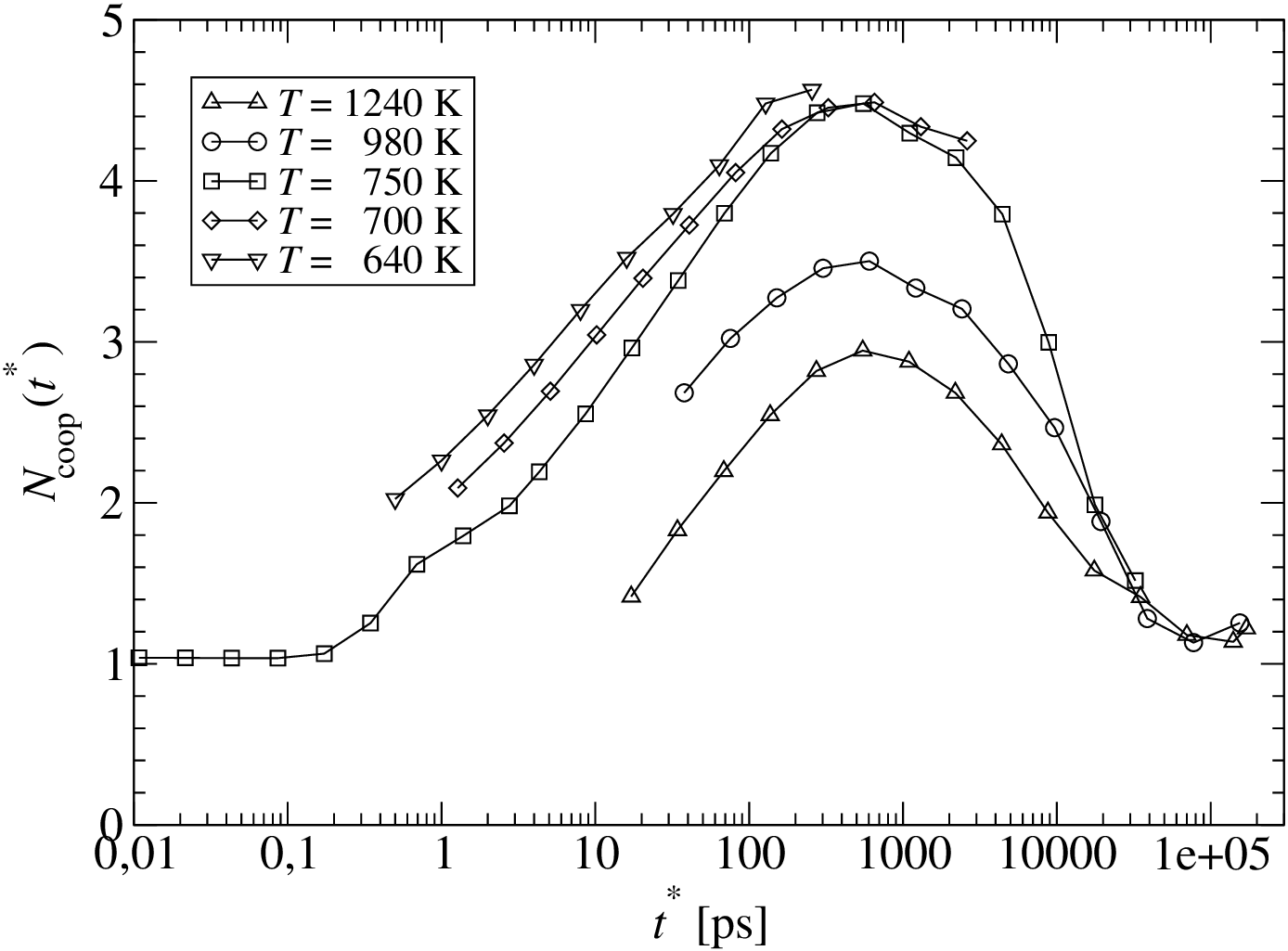}}
\centerline{\bf Fig.\ 8}
\end{figure}

\begin{figure}
\centerline{\epsfxsize=11.5cm\epsffile{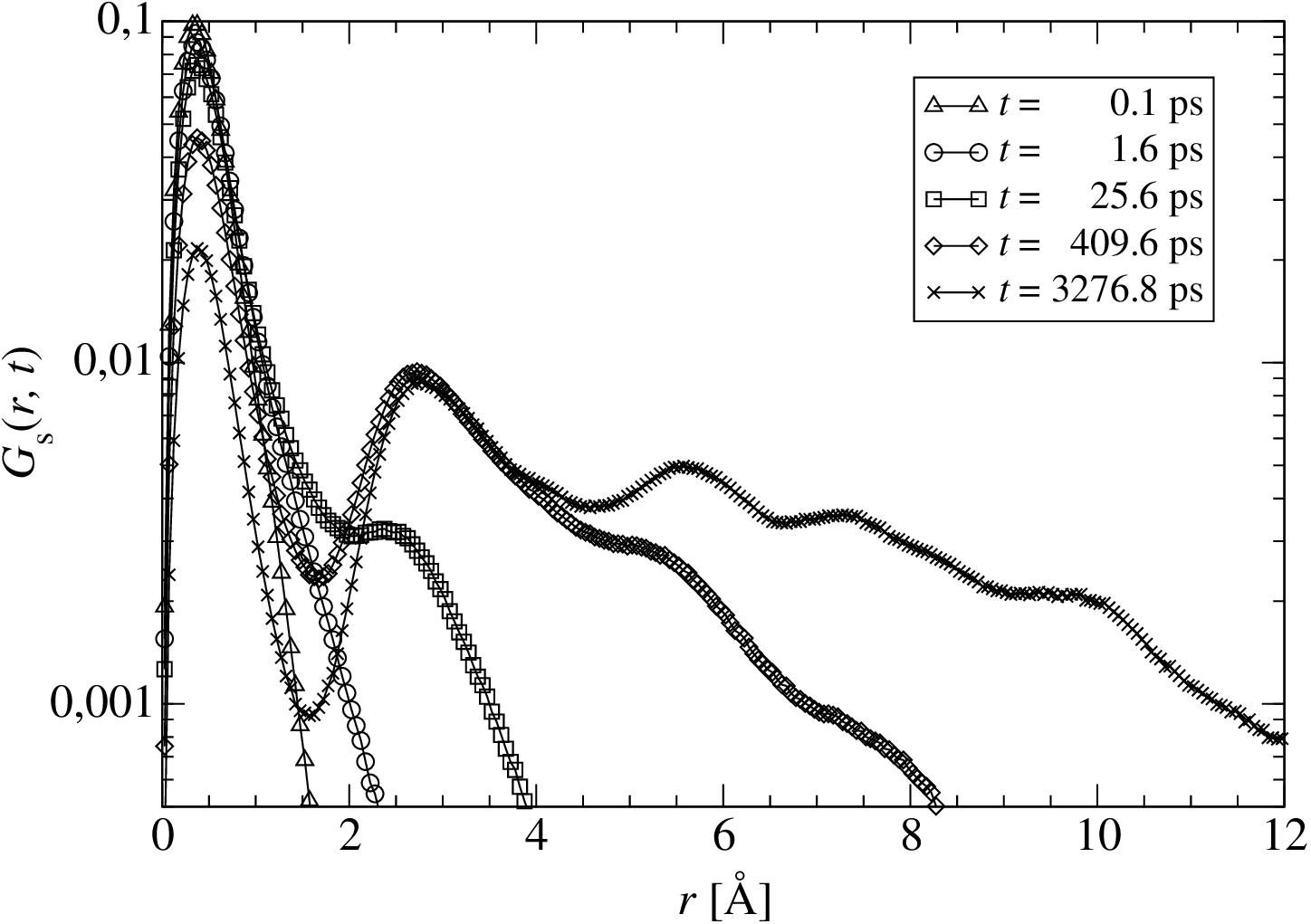}}
\centerline{\bf Fig.\ 9}
\end{figure}
\hfill

\begin{figure}
\centerline{\epsfxsize=11.5cm\epsffile{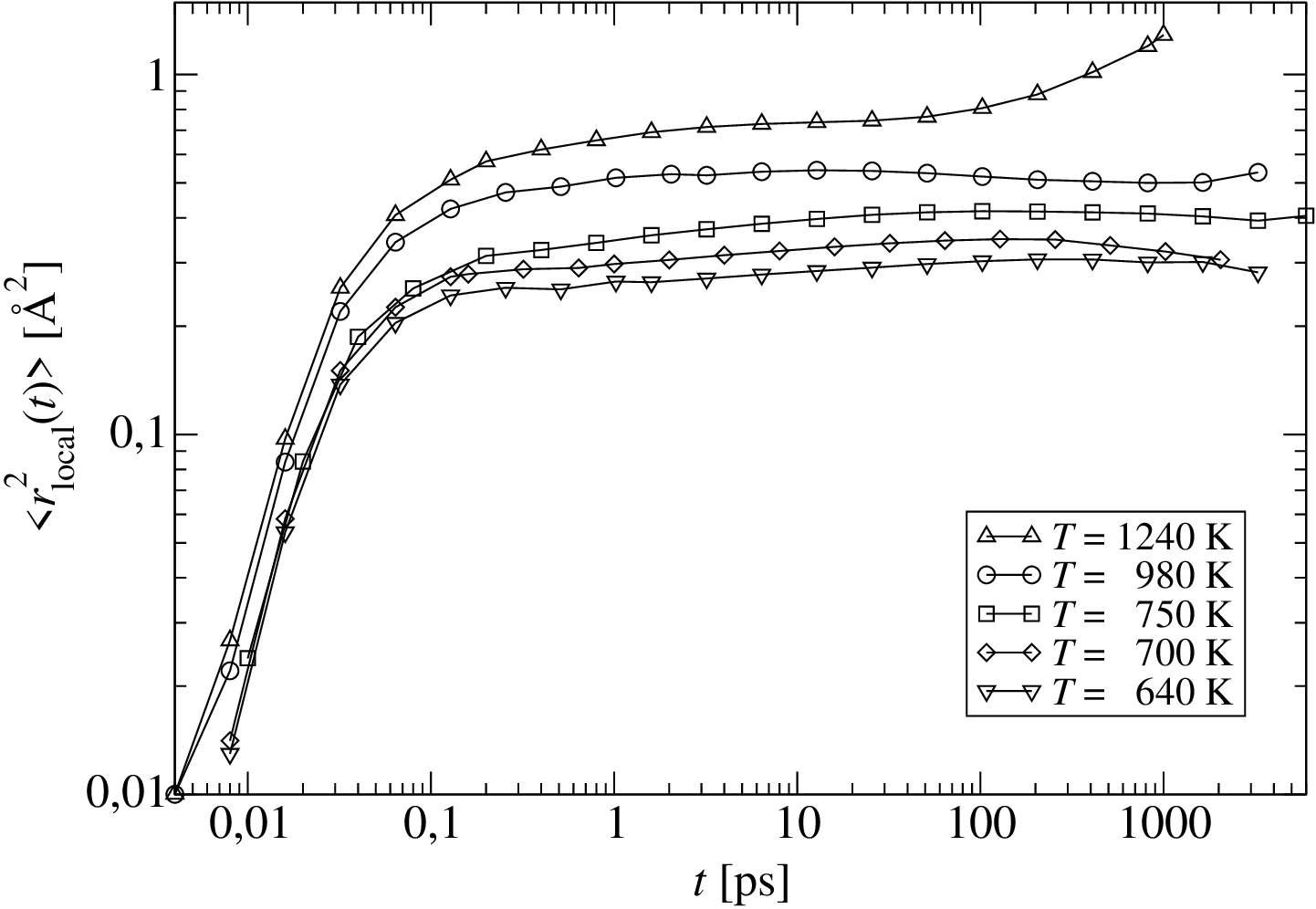}}
\centerline{\bf Fig.\ 10}
\end{figure}

\begin{figure}
\centerline{\epsfxsize=11.5cm\epsffile{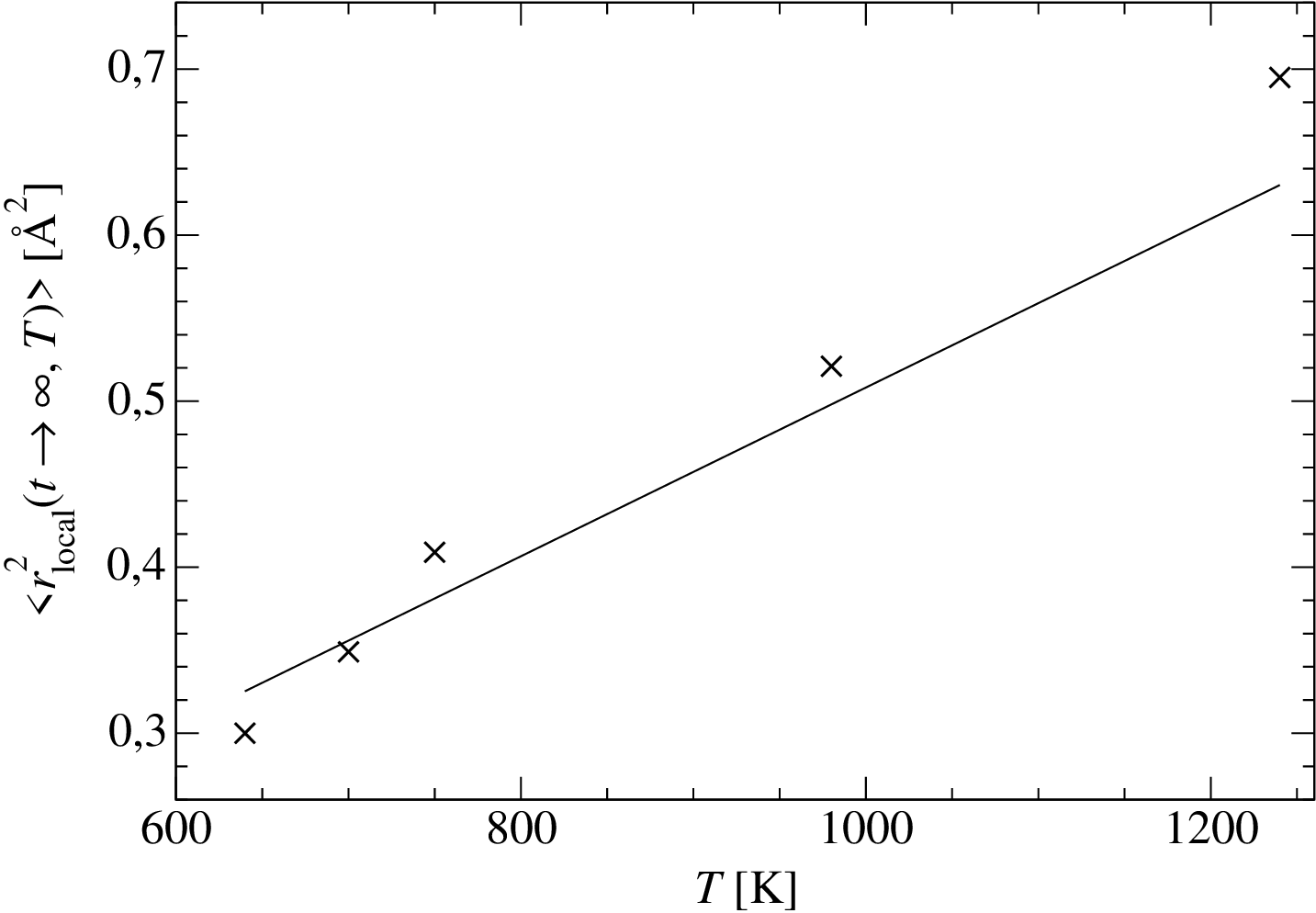}}
\centerline{\bf Fig.\ 11}
\end{figure}
\hfill

\begin{figure}
\centerline{\epsfxsize=11.5cm\epsffile{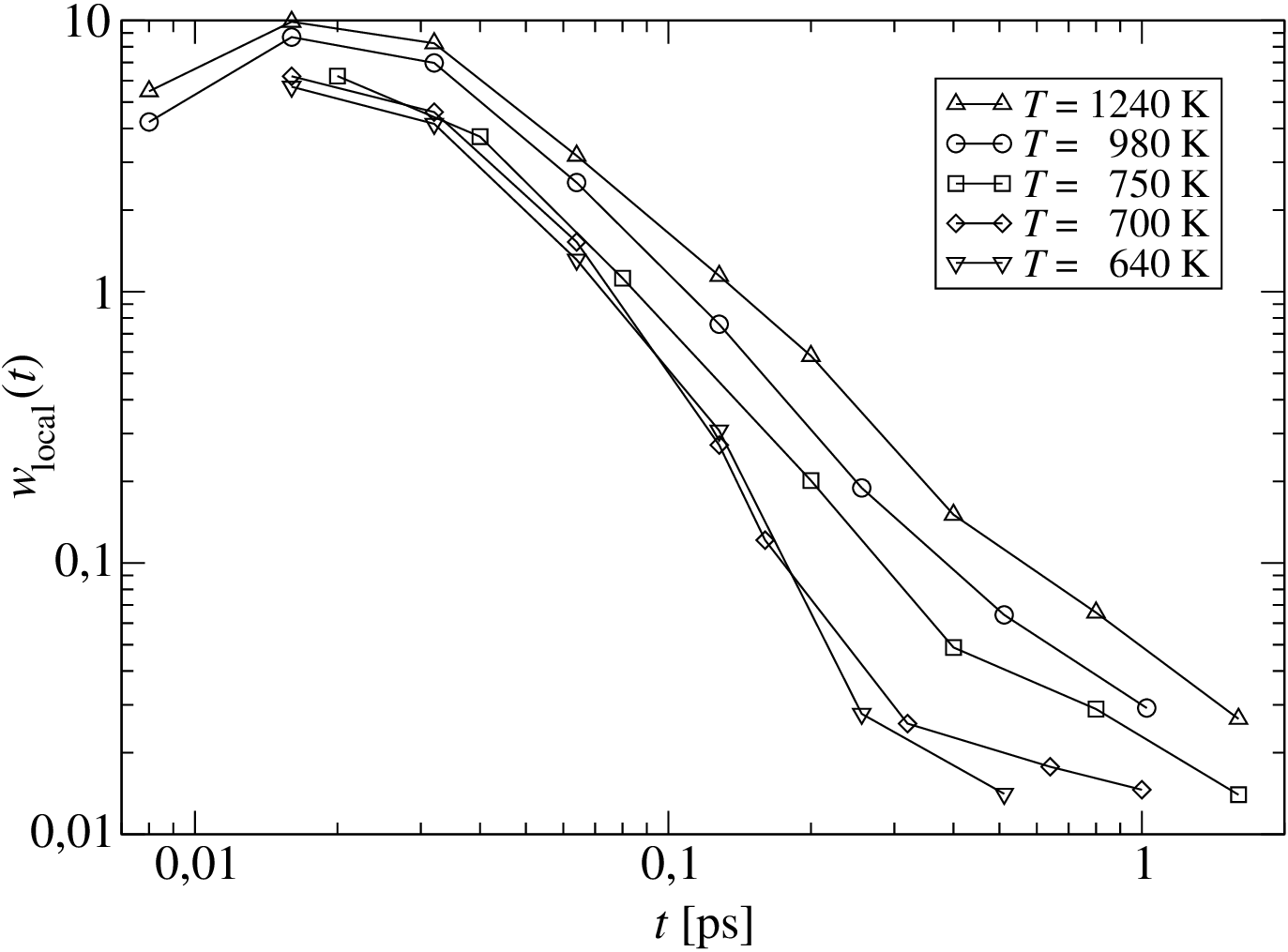}}
\centerline{\bf Fig.\ 12}
\end{figure}

\begin{figure}
\centerline{\epsfxsize=11.5cm\epsffile{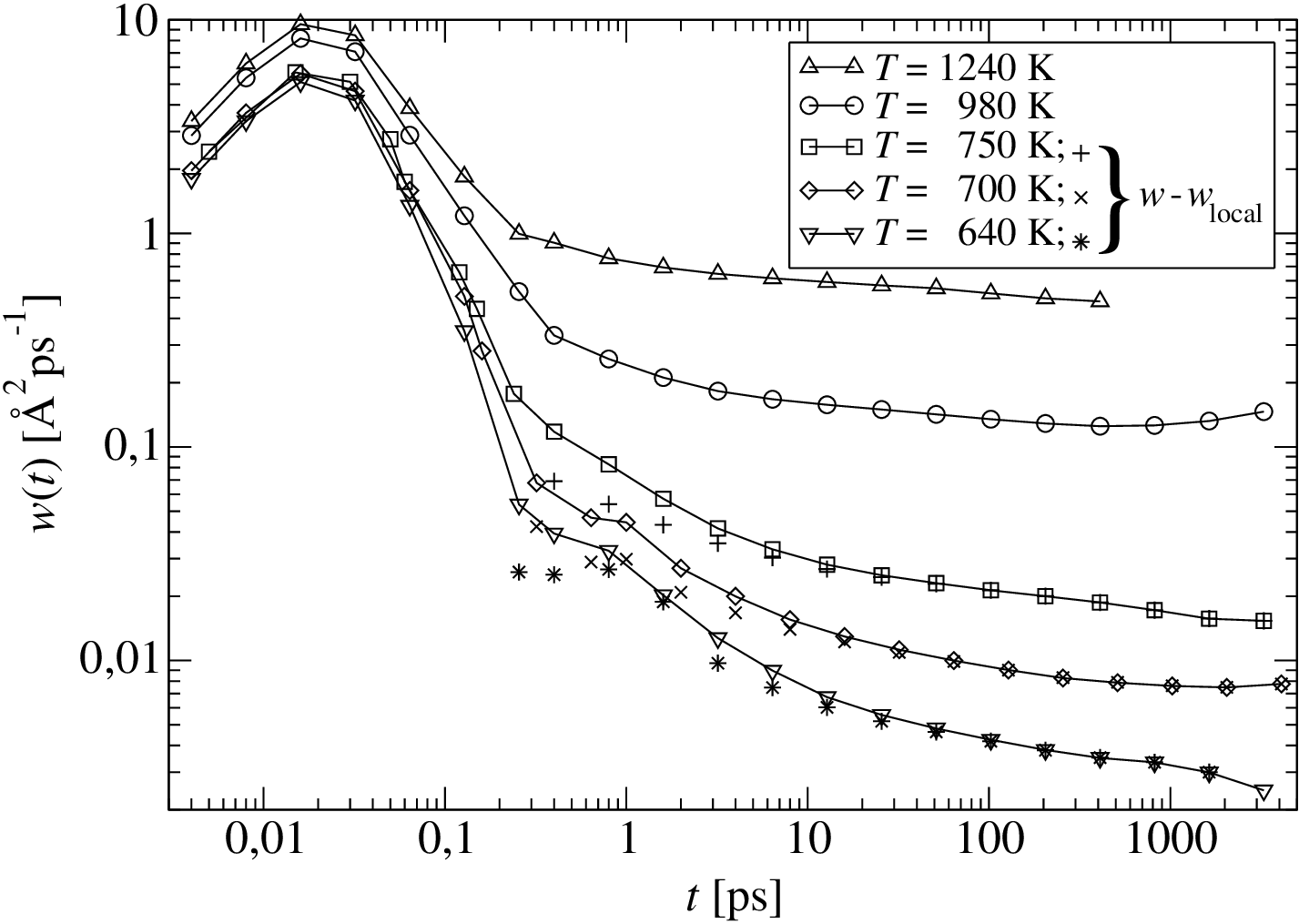}}
\centerline{\bf Fig.\ 13}
\end{figure}
\hfill

\end{document}